\documentstyle[twoside,fleqn,espcrc2]{article}

\newcommand{\be}{\begin{equation}}
\newcommand{\ee}{\end{equation}}

\newcommand{\ba}{\begin{eqnarray}}
\newcommand{\ea}{\end{eqnarray}}

\newcommand{\Tr}{{\rm Tr}}

\title{High--energy scattering amplitudes in QCD:\\
       from Minkowskian to Euclidean space}

\author{E. Meggiolaro\address{Dipartimento di Fisica,
        Universit\`a di Pisa, Via Buonarroti 2, I--56127 Pisa, Italy\\
	E-mail: enrico.meggiolaro@df.unipi.it
}}
\begin{document}

\begin{abstract}

We shall discuss about some analytic properties of the high--energy
parton--parton (and hadron--hadron) scattering amplitudes in gauge theories,
when going from Minkowskian to Euclidean theory, and we shall see how they
can be related to the still unsolved problem of the $s$--dependence of the
total cross--section.

\end{abstract}

\maketitle

\section{Introduction}

\noindent
The parton--parton scattering amplitude, at high squared
energies $s$ in the center of mass and small squared transferred momentum $t$
(that is $s \to \infty$ and $|t| \ll s$, let us say $|t| \le 1~{\rm GeV}^2$),
can be described by the expectation value of two infinite Wilson lines,
running along the classical trajectories of the two colliding particles
\cite{Nachtmann91,Nachtmann97,Meggiolaro96,Meggiolaro01}.

Let us consider, for example, the case of the quark--quark scattering
amplitude.
If one defines the scattering amplitude $T_{fi} = \langle f | \hat{T} |
i \rangle$, between the initial state $| i \rangle$ and the final
state $| f \rangle$, as follows ($\hat{S}$ being the scattering operator)
\be
\langle f | ( \hat{S} - {\bf 1} ) | i \rangle
= i (2\pi)^4 \delta^{(4)} (P_{fin} - P_{in})
~\langle f | \hat{T} | i \rangle ,
\label{Tfi}
\ee
where $P_{in}$ is the initial total four--momentum and $P_{fin}$ is the final
total four--momentum, then, in the center--of--mass reference system (c.m.s.),
taking for example the initial trajectories of the two quarks along the
$x^1$--axis, the high--energy scattering amplitude $T_{fi}$ has the following
form [explicitly indicating the color indices ($i,j, \ldots$)
and the spin indices ($\alpha, \beta, \ldots$) of the quarks]
\cite{Nachtmann91,Nachtmann97,Meggiolaro96,Meggiolaro01}
\ba
\lefteqn{
T_{fi} = \langle \psi_{i\alpha}(p'_1) \psi_{k\gamma}(p'_2) | \hat{T} |
\psi_{j\beta}(p_1) \psi_{l\delta}(p_2) \rangle } \nonumber \\
& & \mathop{\sim}_{s \to \infty}
-i 2s \cdot \delta_{\alpha\beta} \delta_{\gamma\delta}
\cdot {1 \over Z_W^2}
\displaystyle\int d^2 \vec{z}_\perp e^{i \vec{q}_\perp \cdot \vec{z}_\perp}
\nonumber \\
& & \times \langle [ W_{p_1} (z_t) - {\bf 1} ]_{ij}
[ W_{p_2} (0) - {\bf 1} ]_{kl} \rangle,
\label{scatt1}
\ea
where $q = (0,0,\vec{q}_\perp)$, with $t = q^2 = -\vec{q}_\perp^2$, is the
tranferred four--momentum and $z_t = (0,0,\vec{z}_\perp)$, with $\vec{z}_\perp
= (z^2,z^3)$, is the distance between the two trajectories in the
{\it transverse} plane
[the coordinates $(x^0,x^1)$ are often called {\it longitudinal} coordinates].
The expectation
value $\langle f(A) \rangle$ is the average of $f(A)$ in the sense of the
functional integration over the gluon field $A^\mu$ (including also the
determinant of the fermion matrix, i.e., $\det[i\gamma^\mu D_\mu - m_0]$,
where $D^\mu = \partial^\mu + ig A^\mu$ is the covariant derivative and $m_0$
is the {\it bare} quark mass).
The two infinite Wilson lines $W_{p_1} (z_t)$ and $W_{p_2} (0)$ in
Eq. (\ref{scatt1}) are defined as
\ba
W_{p_1} (z_t) =
{\cal T} \exp \left[ -ig \displaystyle\int_{-\infty}^{+\infty}
A_\mu (z_t + p_1 \tilde{\tau}) p_1^\mu d\tilde{\tau} \right]; \nonumber
\ea
\be
W_{p_2} (0) =
{\cal T} \exp \left[ -ig \displaystyle\int_{-\infty}^{+\infty}
A_\mu (p_2 \tilde{\tau}) p_2^\mu d\tilde{\tau} \right],
\label{Wilson0}
\ee
where ${\cal T}$ stands for ``{\it time ordering}'' and $A_\mu = A_\mu^a T^a$;
the four--vectors $p_1 \simeq (E,E,0,0)$ and $p_2 \simeq (E,-E,0,0)$ are the
initial four--momenta of the two quarks [$s = (p_1 + p_2)^2 = 4E^2$].

Finally, $Z_W$ in Eq. (\ref{scatt1}) is the residue at the pole (i.e., for
$p^2 \to m^2$, $m$ being the quark {\it pole} mass) of the unrenormalized
quark propagator, which can be written in the eikonal approximation as
\cite{Nachtmann91,Meggiolaro01}
\ba
\lefteqn{
Z_W \simeq {1 \over N_c} \langle \Tr [ W_{p_1} (z_t) ] \rangle
= {1 \over N_c} \langle \Tr [ W_{p_1} (0) ] \rangle}
\nonumber \\
& & = {1 \over N_c} \langle \Tr [ W_{p_2} (0) ] \rangle,
\label{ZW}
\ea
where $N_c$ is the number of colours.

In a perfectly analogous way, one can also derive the high--energy scattering
amplitude for an elastic process involving two partons, which can be quarks,
antiquarks or gluons \cite{Nachtmann97,Meggiolaro01}.
For an antiquark, one simply has to substitute the Wilson line $W_p(b)$
with its complex conjugate $W^*_p(b)$: this is due to the fact that the
scattering amplitude of an antiquark in the external gluon field $A_\mu$
is equal to the scattering amplitude of a quark in the charge--conjugated
(C--transformed) gluon field $A'_\mu = -A^t_\mu = -A^*_\mu$.
In other words, going from quarks to antiquarks
corresponds just to the change from the fundamental representation $T_a$ of
$SU(N_c)$ to the complex conjugate representation $T'_a = -T^*_a$.
In the same way, going from quarks to gluons
corresponds just to the change from the fundamental representation $T_a$ of
$SU(N_c)$ to the adjoint representation $T^{(adj)}_a$. So, if the parton
is a gluon, one must substitute $W_p(b)$, the Wilson string in the
fundamental representation, with ${\cal V}_k (b)$, the Wilson string
in the adjoint representation [and the renormalization constant $Z_W$
with $Z_{\cal V} = \langle \Tr [{\cal V}_k (0)] \rangle / (N_c^2 - 1)$].

In what follows, to be definite, we shall consider the case of the quark--quark
scattering and we shall deal with the quantity
\ba
\lefteqn{
g_{M (ij,kl)} (s; ~t) \equiv {1 \over Z_W^2}
\displaystyle\int d^2 \vec{z}_\perp e^{i \vec{q}_\perp \cdot \vec{z}_\perp} }
\nonumber \\
& & \times \langle [ W_{p_1} (z_t) - {\bf 1} ]_{ij}
[ W_{p_2} (0) - {\bf 1} ]_{kl} \rangle,
\label{gM}
\ea
in terms of which the scattering amplitude can be written as
\ba
\lefteqn{
T_{fi} = \langle \psi_{i\alpha}(p'_1) \psi_{k\gamma}(p'_2) | \hat{T} |
\psi_{j\beta}(p_1) \psi_{l\delta}(p_2) \rangle }
\nonumber \\
& & \mathop{\sim}_{s \to \infty}
-i 2s \cdot \delta_{\alpha\beta} \delta_{\gamma\delta}
\cdot g_{M (ij,kl)} (s; ~t).
\label{scatt2}
\ea
At first sight, it could appear that the above expression (\ref{gM}) of the
quantity $g_M$ is essentially independent on the center--of--mass energy of
the two quarks and that the $s$--dependence of the scattering amplitude
is all contained in the kinematical factor $2s$ in front of the
integral in Eq. (\ref{scatt1}). This is clearly in contradiction with the
well--known fact that amplitudes in QCD have a very non--trivial
$s$--dependence, whose origin lies in the infrared (IR) divergences typical
of $3 + 1$ dimensional gauge theories.
In more standard perturbative approaches to high--energy QCD, based on the
direct computation of Feynman diagrams in the high--energy limit, these IR
divergences are taken care of by restricting the rapidities of the intermediate
gluons to lie in between those of the two fast quarks (see, e.g.,
\cite{Cheng-Wu-book,Lipatov}).
The classical trajectories of two quarks with a non--zero mass $m$ and a
center--of--mass energy squared $s = 4 E^2$ are related by a finite Lorentz
boost with rapidity parameter $\log (s/m^2)$, so that the size of the rapidity
space for each intermediate gluon grows as $\log s$
and each Feynman diagram acquires an overall
factor proportional to some power of $\log s$, depending on the number of
intermediate gluon propagators.

In the case of the quantity (\ref{gM}), as was first pointed out by Verlinde
and Verlinde in \cite{Verlinde}, the IR singularity is originated by the
fact that the trajectories of the Wilson lines were taken to be lightlike and
therefore have an infinite distance in rapidity space.
One can regularize this infrared problem by giving the Wilson lines a small
timelike component, such that they coincide with the classical trajectories
for quarks with a non--zero mass $m$ (this is equivalent to consider two Wilson
lines forming a certain {\it finite} hyperbolic angle $\chi$ in Minkowskian
space--time; of course, $\chi \to \infty$ when $s \to \infty$),
and, in addition, by letting them end after
some finite proper time $\pm T$ (and eventually letting $T \to \infty$).
Such a regularization of the IR singularities gives rise to an $s$--dependence
of the quantity $g_M$ defined in (\ref{gM}) and, therefore, to a non--trivial
$s$--dependence of the amplitude (\ref{scatt1}), as obtained by ordinary
perturbation theory \cite{Cheng-Wu-book,Lipatov} and as confirmed by the
experiments on hadron--hadron scattering processes. We refer the reader to
Refs.  \cite{Verlinde} and \cite{Meggiolaro97,Meggiolaro98,Meggiolaro02}
for a detailed discussion about this point.

The direct evaluation of the expectation value (\ref{gM}) is a highly
non--trivial matter and it is also strictly connected with the renormalization
properties of Wilson--line operators \cite{Arefeva80,Korchemsky}.
A non--perturbative approach for the calculation of (\ref{gM}) has
been proposed and developed in Refs. \cite{Dosch,Berger},
in the context of the so--called ``stochastic vacuum model''.
In three previous papers \cite{Meggiolaro97,Meggiolaro98,Meggiolaro02} we
proposed and discussed a new approach, which consists in analytically
continuing the scattering amplitude from the Minkowskian to the Euclidean
world, so opening the possibility of studying the scattering amplitude non
perturbatively by well--known and well--established techniques available in
the Euclidean theory (e.g., by means of the formulation of the theory on the
lattice). This approach has been recently adopted in Refs.~\cite{JP1,JP2},
in order to study the high--energy scattering in strongly coupled gauge
theories using the AdS/CFT correspondence, in Ref.~\cite{Shuryak-Zahed},
in order to investigate instanton--induced effects in QCD high--energy
scattering, and also in Ref.~\cite{LLCM}, in the context of the so--called
``loop--loop correlation model'', in which the QCD vacuum is described by
perturbative gluon exchange and the non--perturbative stochastic vacuum
model.

More explicitly, in Refs. \cite{Meggiolaro97,Meggiolaro98} we have given
arguments showing that the expectation value of two {\it infinite} Wilson
lines, forming a certain hyperbolic angle $\chi$ in Minkowskian space--time,
and the expectation value of two {\it infinite} Euclidean Wilson lines,
forming a certain angle $\theta$ in Euclidean four--space, are connected
by an analytic continuation in the angular variables.
This relation of analytic continuation was proved in
Ref. \cite{Meggiolaro97} for an Abelian gauge theory (QED) in the so--called
{\it quenched} approximation and for a non--Abelian gauge theory (QCD) up to
the fourth order in the renormalized coupling constant in perturbation theory;
a general proof was finally given in Ref. \cite{Meggiolaro98}.
The relation of analytic continuation between the amplitudes $g_M (\chi;~t)$
and $g_E (\theta;~t)$, in the Minkowskian and the Euclidean world, was derived
in Refs. \cite{Meggiolaro97,Meggiolaro98} using {\it infinite} Wilson lines,
i.e., directly in the limit $T \to \infty$ and assuming that the amplitudes
were independent on $T$. In other words, the results derived in Refs.
\cite{Meggiolaro97,Meggiolaro98} apply to the cutoff--independent part of the
amplitudes.

On the contrary, in Ref. \cite{Meggiolaro02} we have considered IR--regularized
amplitudes at any $T$ (including also possible divergent pieces when
$T \to \infty$) and, generalizing the results of Ref.  \cite{Meggiolaro98},
we have given the general proof that the expectation value of two
IR--regularized  Wilson lines, forming a certain hyperbolic angle in Minkowskian
space--time, and the expectation value of two IR--regularized Euclidean Wilson
lines, forming a certain angle in Euclidean four--space, are connected by an
analytic continuation in the angular variables and in the IR cutoff $T$.
This result can be used to evaluate the IR--regularized high--energy scattering
amplitude directly in the Euclidean theory, as discussed in Sect. 2.
The conclusions and an outlook are given in Sect. 3.

\section{From Minkowskian to Euclidean theory}

\noindent
Let us consider the following quantity, defined in
Minkowskian space--time:
\ba
g_M (p_1, p_2;~T;~t) = {M (p_1, p_2;~T;~t) \over Z_M(p_1;~T) Z_M(p_2;~T)},
\nonumber
\ea
\ba
\lefteqn{
M (p_1, p_2;~T;~t) =
\displaystyle\int d^2 \vec{z}_\perp e^{i \vec{q}_\perp \cdot \vec{z}_\perp} }
\nonumber \\
& & \times \langle [ W^{(T)}_{p_1} (z_t) - {\bf 1} ]_{ij}
[ W^{(T)}_{p_2} (0) - {\bf 1} ]_{kl} \rangle,
\label{M}
\ea
where $z_t = (0,0,\vec{z}_\perp)$ and $q = (0,0,\vec{q}_\perp)$, so that
$t = -\vec{q}_\perp^2 = q^2$. The Minkowskian four--momenta $p_1$ and $p_2$
are arbitrary four--vectors lying in the longitudinal plane $(x^0,x^1)$
[so that $\vec{p}_{1\perp} = \vec{p}_{2\perp} = \vec{0}_\perp$]
and define the trajectories of the two IR--regularized Wilson lines
$W^{(T)}_{p_1}$ and $W^{(T)}_{p_2}$:
\ba
W^{(T)}_{p_1} (z_t) \equiv
{\cal T} \exp \left[ -ig \displaystyle\int_{-T}^{+T}
A_\mu (z_t + {p_1 \over m} \tau) {p_1^\mu \over m} d\tau \right];
\nonumber
\ea
\be
W^{(T)}_{p_2} (0) \equiv
{\cal T} \exp \left[ -ig \displaystyle\int_{-T}^{+T}
A_\mu ({p_2 \over m} \tau) {p_2^\mu \over m} d\tau \right].
\label{Wilson1}
\ee
$A_\mu = A_\mu^a T^a$ and $m$ is the quark {\it pole} mass.
$T$ is our IR cutoff. \\
$Z_M(p;~T)$ in Eq. (\ref{M}) is defined as ($N_c$ being the number of colours)
\ba
\lefteqn{
Z_M(p;~T) \equiv {1 \over N_c} \langle \Tr [ W^{(T)}_{p} (z_t) ] \rangle }
\nonumber \\
& & = {1 \over N_c} \langle \Tr [ W^{(T)}_{p} (0) ] \rangle.
\label{ZM}
\ea
(The last equality comes from the space--time translation invariance.)
This is a sort of Wilson--line's renormalization constant:
as shown in Ref. \cite{Meggiolaro01}, $Z_M(p~;T \to \infty)$ is the residue at
the pole (i.e., for $p^2 \to m^2$) of the unrenormalized quark propagator,
in the eikonal approximation.

In an analogous way, we can consider the following quantity, defined
in Euclidean four--space:
\ba
g_E (p_{1E}, p_{2E};~T;~t) = {E (p_{1E}, p_{2E};~T;~t) \over
Z_E(p_{1E};~T) Z_E(p_{2E};~T)}, \nonumber
\ea
\ba
\lefteqn{
E (p_{1E}, p_{2E};~T;~t) =
\displaystyle\int d^2 \vec{z}_\perp e^{i \vec{q}_\perp \cdot \vec{z}_\perp} }
\nonumber \\
& & \times \langle [ \tilde{W}^{(T)}_{p_{1E}} (z_{t E}) - {\bf 1} ]_{ij}
[ \tilde{W}^{(T)}_{p_{2E}} (0) - {\bf 1} ]_{kl} \rangle_E,
\label{E}
\ea
where $z_{t E} = (0, \vec{z}_\perp, 0)$ and
$q_E = (0, \vec{q}_\perp, 0)$, so that: $t = -\vec{q}^2_\perp = -q_E^2$.
The expectation value $\langle \ldots \rangle_E$ must be intended now as a
functional integration with respect to the gauge variable $A^{(E)}_\mu =
A^{(E)a}_\mu T^a$ in the Euclidean theory.
The Euclidean four--momenta $p_{1E}$ and $p_{2E}$ are arbitrary four--vectors
lying in the plane $(x_1,x_4)$ [so that $\vec{p}_{1E\perp} = \vec{p}_{2E\perp}
= \vec{0}_\perp$] and define the trajectories of the two IR--regularized
Euclidean Wilson lines $\tilde{W}^{(T)}_{p_{1E}}$ and
$\tilde{W}^{(T)}_{p_{2E}}$:
\ba
\lefteqn{
\tilde{W}^{(T)}_{p_{1E}} (z_{t E}) \equiv }
\nonumber \\
& & {\cal T} \exp \left[ -ig \displaystyle\int_{-T}^{+T}
A^{(E)}_{ \mu} (z_{t E} + {p_{1E} \over m} \tau) {p_{1E \mu} \over m} d\tau
\right]; \nonumber
\ea
\ba
\lefteqn{
\tilde{W}^{(T)}_{p_{2E}} (0) \equiv }
\nonumber \\
& & {\cal T} \exp \left[ -ig \displaystyle\int_{-T}^{+T}
A^{(E)}_{ \mu} ({p_{2E} \over m} \tau) {p_{2E \mu} \over m} d\tau \right].
\label{Wilson2}
\ea
$Z_E(p_E;~T)$ in Eq. (\ref{E}) is defined analogously to $Z_M(p;~T)$
in Eq. (\ref{ZM}):
\ba
\lefteqn{
Z_E(p_E;~T) \equiv {1 \over N_c} \langle \Tr [ \tilde{W}^{(T)}_{p_E}
(z_{t E}) ] \rangle_E }
\nonumber \\
& & = {1 \over N_c} \langle \Tr [ \tilde{W}^{(T)}_{p_E} (0) ] \rangle_E.
\label{ZE}
\ea
(The last equality comes from the translation invariance in Euclidean
four--space.)

Since we finally want to obtain the expression (\ref{scatt1}) of the
scattering amplitude in the c.m.s. of the two quarks, taking their
initial trajectories along the $x^1$--axis, we {\it choose} $p_1$ and $p_2$
to be the four--momenta of the two particles with mass $m$, moving with speed
$\beta$ and $-\beta$ along the $x^1$--direction, i.e.,
\ba
p_1 &=& E (1,\beta,0,0), \nonumber \\
p_2 &=& E (1,-\beta,0,0),
\label{p12}
\ea
where $E = m / \sqrt{1 - \beta^2}$ (in units with $c=1$) is the
energy of each particle (so that: $s = 4E^2$).
We now introduce the hyperbolic angle $\psi$ [in the plane $(x^0,x^1)$]
of the trajectory of $W^{(T)}_{p_1}$: it is given by $\beta = \tanh \psi$.
We can give the explicit form of the Minkowskian four--vectors $p_1$
and $p_2$ in terms of the hyperbolic angle $\psi$:
\ba
p_1 &=& m (\cosh \psi,\sinh \psi,0,0), \nonumber \\
p_2 &=& m (\cosh \psi,-\sinh \psi,0,0).
\label{p12M}
\ea
Clearly, $p_1^2 = p_2^2 = m^2$ and
\be
p_1 \cdot p_2 = m^2 \cosh (2\psi) = m^2 \cosh \chi,
\label{chi}
\ee
where $\chi = 2\psi$ is the hyperbolic angle [in the plane $(x^0,x^1)$]
between the two trajectories of $W^{(T)}_{p_1}$ and $W^{(T)}_{p_2}$.

Analogously, in the Euclidean theory we {\it choose}
a reference frame in which the spatial vectors $\vec{p}_{1E}$ and
$\vec{p}_{2E} = -\vec{p}_{1E}$ are along the $x_1$--direction and,
moreover, $p_{1E}^2 = p_{2E}^2 = m^2$; that is:
\ba
p_{1E} &=& m (\sin \phi, 0, 0, \cos \phi ); \nonumber \\
p_{2E} &=& m (-\sin \phi, 0, 0, \cos \phi ),
\label{p12E}
\ea
where $\phi$ is the angle formed by each trajectory with the $x_4$--axis.
The value of $\phi$ is between $0$ and $\pi / 2$, so that the angle
$\theta = 2 \phi$ between the two Euclidean trajectories
$\tilde{W}^{(T)}_{p_{1E}}$ and $\tilde{W}^{(T)}_{p_{2E}}$
lies in the range $[0,\pi]$: it is always possible to make
such a choice by virtue of the $O(4)$ symmetry of the Euclidean theory.
From (\ref{p12E}) we derive that:
\be
p_{1E} \cdot p_{2E} = m^2 \cos \theta.
\label{theta}
\ee
[A short remark about the notation: we have denoted everywhere the
scalar product by a ``$\cdot$'', both in the Minkowskian and the Euclidean
world. Of course, when $A$ and $B$ are Minkowskian four--vectors, then
$A \cdot B = A^\mu B_\mu = A^0 B^0 - \vec{A} \cdot \vec{B}$; while, if
$A_E$ and $B_E$ are Euclidean four--vectors, then
$A_E \cdot B_E = A_{E \mu} B_{E \mu} = \vec{A}_E \cdot \vec{B}_E +
A_{E 4} B_{E 4}$.]

It has been proved in Ref. \cite{Meggiolaro02} that, if we denote with
$M(\chi;~T;~t)$ the value of $M(p_1, p_2;~T;~t)$
for $p_1$ and $p_2$ given by Eq. (\ref{p12M}) and we also denote with
$E(\theta;~T;~t)$ the value of $E (p_{1E}, p_{2E};~T;~t)$
for $p_{1E}$ and $p_{2E}$ given by Eq. (\ref{p12E}),
the following relation holds (reminding that
$\phi = \theta/2$ and $\psi = \chi/2$):
\be
E (\theta;~T;~t) = M (\chi \to i\theta;~T \to -iT;~t).
\label{EMres4}
\ee
Let us consider, now, the Wilson--line's renormalization constants $Z_M(p;~T)$
in the Minkowskian theory and $Z_E(p_E;~T)$ in the Euclidean theory,
defined by Eqs. (\ref{ZM}) and (\ref{ZE}) respectively.

From the definition (\ref{ZM}), $Z_M (p;~T)$, considered as a function of a
general four--vector $p$, is a scalar function constructed
with the only four--vector $u \equiv p/m$.
In a perfectly analogous way, from the definition (\ref{ZE}) in the Euclidean
case, $Z_E (p_E;~T)$, considered as a function of a general Euclidean
four--vector $p_E$, is a scalar function constructed with the only Euclidean
four--vector $u_E \equiv p_E/m$.
It has been proved in Ref. \cite{Meggiolaro02} that, if we denote with
$Z_W(T)$ the value of $Z_M(p_1;~T)$ or $Z_M(p_2;~T)$
for $p_1$ and $p_2$ given by Eq. (\ref{p12M}) and we also denote with
$Z_{W E}(T)$ the value of $Z_E(p_{1E};~T)$ or $Z_E(p_{2E};~T)$
for $p_{1E}$ and $p_{2E}$ given by Eq. (\ref{p12E}),
the following relation holds:
\be
Z_{W E}(T) = Z_W(-iT).
\label{ZEMres4}
\ee
Combining this identity with Eq. (\ref{EMres4}), we find that the Minkowskian
and the Euclidean amplitudes, defined by Eqs. (\ref{M}) and (\ref{E}),
with $p_1$ and $p_2$ given by Eq. (\ref{p12M}) and $p_{1E}$
and $p_{2E}$ given by Eq. (\ref{p12E}), i.e.,
\ba
g_M (\chi;~T;~t) &\equiv& {M (\chi;~T;~t) \over [Z_W(T)]^2},
\nonumber \\
g_E (\theta;~T;~t) &\equiv& {E (\theta;~T;~t) \over [Z_{W E}(T)]^2},
\label{gM-gE}
\ea
are connected by the following relation \cite{Meggiolaro02}:
\ba
g_E (\theta;~T;~t) = g_M (\chi \to i\theta;~T \to -iT;~t);
\nonumber \\
g_M (\chi;~T;~t) = g_E (\theta \to -i\chi;~T \to iT;~t).
\label{final}
\ea
The relation (\ref{final}) of analytic continuation has been derived for a
non--Abelian gauge theory with gauge group $SU(N_c)$. It is clear, however,
from the derivation fully reported in Ref. \cite{Meggiolaro02}, that the same
result is valid also for an Abelian gauge theory (QED).

Moreover, even if the result (\ref{final}) has been explicitly derived for the
case of the quark--quark scattering, it is immediately generalized to the more
generale case of the parton--parton scattering, where each parton can be a
quark, an antiquark or a gluon. In fact, as explained in the Introduction,
one simply has to use a proper Wilson line for each parton: $W_p (b)$, the
Wilson string in the fundamental representation $T^a$, for a quark;
$W^*_p (b)$, the Wilson string in the complex conjugate representation
$T'_a = -T^*_a$, for an antiquark; and ${\cal V}_k (b)$, the Wilson string
in the adjoint representation $T^{(adj)}_a$, for a gluon.
The proof leading to Eq. (\ref{final}) is then repeated step by step, after
properly modifying the definitions (\ref{Wilson1}) and (\ref{Wilson2})
of the Wilson lines.
[If the parton is a gluon, one must substitute the quark mass $m$ appearing
in all previous formulae with an arbitrarily small mass $\mu \to 0$. The
IR cutoff appears in all expressions in the form of the ratio $T/\mu$ for
a gluon and $T/m$ for a quark/antiquark.]

The relation (\ref{final}), originally derived in Ref. \cite{Meggiolaro02},
completely generalizes the results of Ref.
\cite{Meggiolaro98}, where we derived a relation of analytic continuation
between the amplitudes $g_M (\chi;~t)$ and $g_E (\theta;~t)$, in the
Minkowskian and the Euclidean world, using {\it infinite} Wilson lines, i.e.,
directly in the limit $T \to \infty$ and assuming that the amplitudes
were independent on $T$. In other words, we can claim that the results
of Ref. \cite{Meggiolaro98} apply to the cutoff--independent part of the
amplitudes, while Eq. (\ref{final}) is a relation of analytic continuation
between IR--regularized amplitudes at any $T$.

The result (\ref{final}) can be used to evaluate
the IR--regularized high--energy parton--parton scattering amplitude
directly in the Euclidean theory. In fact, the IR--regularized
high--energy scattering amplitude is given (e.g., for the case of the
quark--quark scattering) by
\ba
\lefteqn{
T_{fi} = \langle \psi_{i\alpha}(p'_1) \psi_{k\gamma}(p'_2) | \hat{T} |
\psi_{j\beta}(p_1) \psi_{l\delta}(p_2) \rangle
\mathop{\sim}_{s \to \infty} }
\nonumber \\
& & -i 2s \cdot \delta_{\alpha\beta} \delta_{\gamma\delta}
\cdot g_M (\chi \to \infty;~T \to \infty;~t),
\label{scatt4}
\ea
where the quantity $g_M (\chi;~T;~t)$, defined by Eq. (\ref{M}), is
essentially a correlation function of two IR--regularized Wilson lines
forming a certain hyperbolic angle $\chi$ in Minkowskian space--time.
For deriving the dependence on $s$ one exploits the fact that the hyperbolic
angle $\chi$ is a function of $s$. In fact, from $s = 4E^2$,
$E = m/\sqrt{1 - \beta^2}$, and $\beta = \tanh (\chi/2)$ [see Eqs. (\ref{p12}),
(\ref{p12M}) and (\ref{chi})], one immediately finds that:
\be
s = 2 m^2 ( \cosh \chi + 1 ).
\label{s-chi}
\ee
Therefore, in the high--energy limit $s \to \infty$ (or $\chi \to \infty$,
i.e., $\beta \to 1$), the hyperbolic angle $\chi$ is essentially equal
to the logarithm of $s/m^2$ (for a non--zero quark mass $m$):
\be
\chi \mathop{\sim}_{s \to \infty} \log \left( {s \over m^2} \right).
\label{logs}
\ee
The quantity $g_M (\chi;~T;~t)$ is linked to the corresponding Euclidean
quantity $g_E (\theta;~T;~t)$, defined by Eq. (\ref{E}),
by the analytic continuation (\ref{final}) in the angular variables and in
the IR cutoff $T$.
Therefore, one can start by evaluating $g_E (\theta;~T;~t)$, which is
essentially a correlation function of two IR--regularized Wilson lines
forming a certain angle $\theta$ in Euclidean four--space, and then one can
continue this quantity into Minkowskian space--time by rotating the Euclidean
angular variable clockwise, $\theta \to -i \chi$, and the IR cutoff (Euclidean
proper time) anticlockwise, $T \to iT$: in such a way one reconstructs the
Minkowskian quantity $g_M (\chi;~T;~t)$.
As was pointed out in \cite{JP2}, one should note that {\it a priori} there is
an ambiguity in making such an analytic continuation, depending on the
precise choice of the path. This phenomenon does not
appear when the Euclidean correlation function $g_E(\theta;~T;t)$ has only
simple poles in the complex $\theta$--plane, but in some cases the analyticity
structure can contain branch cuts in the complex plane, which must be taken
into account: we refer the reader to Ref. \cite{JP2} for a full
discussion about this point.

\section{From Wilson lines to Wilson loops}

\noindent
We want to conclude by making a remark about the problem of the IR
divergences which appear in the high--energy scattering amplitudes.

A well--known feature of the parton--parton scattering amplitude is its
IR divergence, which, as we have already said in the Introduction, is typical
of $3 + 1$ dimensional gauge theories and which, in our formulation,
manifests itself in the IR singularity of the correlation
function of two Wilson lines  when $T \to \infty$.
In many cases these IR divergences can be factorized out.

As suggested in Ref. \cite{JP2}, an alternative way to eliminate this
cutoff dependence is to consider an IR--finite physical quantity, like
the scattering amplitude of two colourless states in gauge theories,
e.g., two $q \bar{q}$ meson states.
It was shown in Ref. \cite{Nachtmann97} that the high--energy meson--meson
scattering amplitude can be approximately reconstructed by first evaluating,
in the eikonal approximation, the scattering amplitude of two $q \bar{q}$
pairs, of given transverse sizes $\vec{R}_{1\perp}$ and $\vec{R}_{2\perp}$
respectively, and then folding this amplitude with two proper
wave functions $\omega_1 (\vec{R}_{1\perp})$ and $\omega_2 (\vec{R}_{2\perp})$
describing the two interacting mesons. It turns out that the high--energy
scattering amplitude of two $q \bar{q}$ pairs of transverse sizes
$\vec{R}_{1\perp}$ and $\vec{R}_{2\perp}$, and impact--parameter distance
$\vec{z}_\perp$, is governed by the correlation function of two Wilson loops
${\cal W}_1$ and ${\cal W}_2$, which follow the classical straight lines for
quark (antiquark) trajectories \cite{Nachtmann97,Dosch}:
\ba
{\cal W}^{(T)}_1 (\vec{z}_\perp,\vec{R}_{1\perp}) \to
X_{\pm 1}^\mu(\tau) = z_t^\mu + {p_1^\mu \over m} \tau
\pm {R_{1t}^\mu \over 2}; \nonumber
\ea
\be
{\cal W}^{(T)}_2 (\vec{0}_\perp,\vec{R}_{2\perp}) \to
X_{\pm 2}^\mu(\tau) = {p_2^\mu \over m} \tau \pm {R_{2t}^\mu \over 2},
\label{loops}
\ee
[where, as usual, $z_t = (0,0,\vec{z}_\perp)$ and also
$R_{1t} = (0,0,\vec{R}_{1\perp})$ and $R_{2t} = (0,0,\vec{R}_{2\perp})$]
and close at proper times $\tau = \pm T$. \\
The same analytic continuation (\ref{final}), that has been derived for the
case of Wilson lines, is, of course, expected to apply also
to the Wilson--loop correlators: the proof can be repeated going
essentially through the same steps (see Ref. \cite{Meggiolaro02}), after
adapting the definitions (\ref{Wilson1}) and (\ref{Wilson2})
from the case of Wilson lines to the case of Wilson loops.
However, in this case the cutoff dependence on $T$ is expected to be removed
together with the related IR divergence which was present for the
case of Wilson lines.
As an example to illustrate these considerations, let us consider the simple
case of {\it quenched} QED. In this case, the calculation of the correlator
of the two Wilson loops given in Eq. (\ref{loops}) can be performed
explicitly and one finds the following result, in the Minkowskian space--time:
\ba
\lefteqn{
{\langle {\cal W}^{(T)}_1 (\vec{z}_\perp,\vec{R}_{1\perp})
{\cal W}^{(T)}_2 (\vec{0}_\perp,\vec{R}_{2\perp}) \rangle \over
\langle {\cal W}^{(T)}_1 (\vec{z}_\perp,\vec{R}_{1\perp}) \rangle
\langle {\cal W}^{(T)}_2 (\vec{0}_\perp,\vec{R}_{2\perp}) \rangle}
\mathop\simeq_{T \to \infty} }
\nonumber \\
& & \exp \Biggl[ -i 4e^2 {\rm cotgh}\chi \displaystyle\int {d^2 \vec{k}_\perp
\over (2\pi)^2} {e^{-i\vec{k}_\perp \cdot \vec{z}_\perp} \over \vec{k}^2_\perp}
\Biggr.  \nonumber \\
& & \Biggl. \times \sin \left( {\vec{k}_\perp \cdot \vec{R}_{1\perp} \over 2}
\right) \sin \left( {\vec{k}_\perp \cdot \vec{R}_{2\perp} \over 2} \right)
\Biggr],
\label{corr-loop-M}
\ea
where $\chi$ is, as usual, the hyperbolic angle between the two Wilson loops.
The analogous calculation in the Euclidean space gives the following result:
\ba
\lefteqn{
{\langle \tilde{\cal W}^{(T)}_{1E} (\vec{z}_\perp,\vec{R}_{1\perp})
\tilde{\cal W}^{(T)}_{2E} (\vec{0}_\perp,\vec{R}_{2\perp}) \rangle_E \over
\langle \tilde{\cal W}^{(T)}_{1E} (\vec{z}_\perp,\vec{R}_{1\perp}) \rangle_E
\langle \tilde{\cal W}^{(T)}_{2E} (\vec{0}_\perp,\vec{R}_{2\perp}) \rangle_E}
\mathop\simeq_{T \to \infty} }
\nonumber \\
& & \exp \Biggl[ -4e^2 {\rm cotg}\theta \displaystyle\int {d^2 \vec{k}_\perp
\over (2\pi)^2} {e^{-i\vec{k}_\perp \cdot \vec{z}_\perp} \over \vec{k}^2_\perp}
\Biggr. \nonumber \\
& & \Biggl. \times \sin \left( {\vec{k}_\perp \cdot \vec{R}_{1\perp} \over 2}
\right) \sin \left( {\vec{k}_\perp \cdot \vec{R}_{2\perp} \over 2} \right)
\Biggr],
\label{corr-loop-E}
\ea
where $\theta$ is the angle between the two Euclidean Wilson loops.
One can see explicitly that the two quantities (\ref{corr-loop-M}) and
(\ref{corr-loop-E}) are indeed IR finite (when $T \to \infty$) and are
connected by the usual analytic continuation in the angular variables
($\chi \to i\theta$).

In our opinion, the high--energy scattering
problem could be directly investigated on the lattice using this
Wilson--loop formulation.
A further advantage of the Wilson--loop formulation, which makes it suitable
to be studied on the lattice, is that, contrary to the Wilson--line
formulation, it is manifestly gauge--invariant.
(In the case of the parton--parton scattering amplitude, gauge invariance can
be restored, at least for the {\it diffractive}, i.e., no--colour--exchange,
part proportional to $\langle \Tr [W_{p_1}(z_t) - {\bf 1}] \Tr [W_{p_2}(0)
- {\bf 1}] \rangle$, by requiring that the gauge transformations at both ends
of the Wilson lines are the same \cite{Nachtmann91,Verlinde}.) \\
A considerable progress is expected along this line in the near future.

\end{document}